\magnification 1200

%
%
\newdimen\FigSize       \FigSize=.9\hsize 
%
\newskip\abovefigskip   \newskip\belowfigskip
\gdef\epsfig#1;#2;{\par\vskip\abovefigskip\penalty -500
   {\everypar={}\epsfxsize=#1\nd
    \centerline{\epsfbox{#2}}}%
    \vskip\belowfigskip}%
%
\newskip\figtitleskip
\gdef\tepsfig#1;#2;#3{\par\vskip\abovefigskip\penalty -500
   {\everypar={}\epsfxsize=#1\nd
    \vbox
      {\centerline{\epsfbox{#2}}\vskip\figtitleskip
       \centerline{\figtitlefont#3}}}%
    \vskip\belowfigskip}%
%
\newcount\FigNr \global\FigNr=0
\gdef\nepsfig#1;#2;#3{\global\advance\FigNr by 1
   \tepsfig#1;#2;{Figure\space\the\FigNr.\space#3}}%
%
%
%
\gdef\ipsfig#1;#2;{
   \midinsert{\everypar={}\epsfxsize=#1\nd
              \centerline{\epsfbox{#2}}}%
   \endinsert}%
%
\gdef\tipsfig#1;#2;#3{\midinsert
   {\everypar={}\epsfxsize=#1\nd
    \vbox{\centerline{\epsfbox{#2}}%
          \vskip\figtitleskip
          \centerline{\figtitlefont#3}}}\endinsert}%
%
\gdef\nipsfig#1;#2;#3{\global\advance\FigNr by1%
  \tipsfig#1;#2;{Figure\space\the\FigNr.\space#3}}%
\newread\epsffilein    
\newif\ifepsffileok    
\newif\ifepsfbbfound   
\newif\ifepsfverbose   
\newdimen\epsfxsize    
\newdimen\epsfysize    
\newdimen\epsftsize    
\newdimen\epsfrsize    
\newdimen\epsftmp      
\newdimen\pspoints     
\pspoints=1bp          
\epsfxsize=0pt         
\epsfysize=0pt         
\def\epsfbox#1{\global\def\epsfllx{72}\global\def\epsflly{72}%
   \global\def\epsfurx{540}\global\def\epsfury{720}%
   \def\lbracket{[}\def\testit{#1}\ifx\testit\lbracket
   \let\next=\epsfgetlitbb\else\let\next=\epsfnormal\fi\next{#1}}%
\def\epsfgetlitbb#1#2 #3 #4 #5]#6{\epsfgrab #2 #3 #4 #5 .\\%
   \epsfsetgraph{#6}}%
\def\epsfnormal#1{\epsfgetbb{#1}\epsfsetgraph{#1}}%
\def\epsfgetbb#1{%
%
%
\openin\epsffilein=#1
\ifeof\epsffilein\errmessage{I couldn't open #1, will ignore it}\else
%
%
   {\epsffileoktrue \chardef\other=12
    \def\do##1{\catcode`##1=\other}\dospecials \catcode`\ =10
    \loop
       \read\epsffilein to \epsffileline
       \ifeof\epsffilein\epsffileokfalse\else
%
%
          \expandafter\epsfaux\epsffileline:. \\%
       \fi
   \ifepsffileok\repeat
   \ifepsfbbfound\else
    \ifepsfverbose\message{No bounding box comment in #1; using
defaults}\fi\fi
   }\closein\epsffilein\fi}%
%
%
\def\epsfsetgraph#1{%
   \epsfrsize=\epsfury\pspoints
   \advance\epsfrsize by-\epsflly\pspoints
   \epsftsize=\epsfurx\pspoints
   \advance\epsftsize by-\epsfllx\pspoints
%
%
   \epsfxsize\epsfsize\epsftsize\epsfrsize
   \ifnum\epsfxsize=0 \ifnum\epsfysize=0
      \epsfxsize=\epsftsize \epsfysize=\epsfrsize
%
%
     \else\epsftmp=\epsftsize \divide\epsftmp\epsfrsize
       \epsfxsize=\epsfysize \multiply\epsfxsize\epsftmp
       \multiply\epsftmp\epsfrsize \advance\epsftsize-\epsftmp
       \epsftmp=\epsfysize
       \loop \advance\epsftsize\epsftsize \divide\epsftmp 2
       \ifnum\epsftmp>0
          \ifnum\epsftsize<\epsfrsize\else
             \advance\epsftsize-\epsfrsize \advance\epsfxsize\epsftmp
\fi
       \repeat
     \fi
   \else\epsftmp=\epsfrsize \divide\epsftmp\epsftsize
     \epsfysize=\epsfxsize \multiply\epsfysize\epsftmp
     \multiply\epsftmp\epsftsize \advance\epsfrsize-\epsftmp
     \epsftmp=\epsfxsize
     \loop \advance\epsfrsize\epsfrsize \divide\epsftmp 2
     \ifnum\epsftmp>0
        \ifnum\epsfrsize<\epsftsize\else
           \advance\epsfrsize-\epsftsize \advance\epsfysize\epsftmp \fi
     \repeat
   \fi
%
%
   \ifepsfverbose\message{#1: width=\the\epsfxsize,
height=\the\epsfysize}\fi
   \epsftmp=10\epsfxsize \divide\epsftmp\pspoints
   \vbox to\epsfysize{\vfil\hbox to\epsfxsize{%
      \includegraphics{#1}%
      \hfil}}%
\epsfxsize=0pt\epsfysize=0pt}%
%
%
{\catcode`\%=12
\global\let\epsfpercent=
%
%
\long\def\epsfaux#1#2:#3\\{\ifx#1\epsfpercent
   \def\testit{#2}\ifx\testit\epsfbblit
      \epsfgrab #3 . . . \\%
      \epsffileokfalse
      \global\epsfbbfoundtrue
   \fi\else\ifx#1\par\else\epsffileokfalse\fi\fi}%
%
%
\def\epsfgrab #1 #2 #3 #4 #5\\{%
   \global\def\epsfllx{#1}\ifx\epsfllx\empty
      \epsfgrab #2 #3 #4 #5 .\\\else
   \global\def\epsflly{#2}%
   \global\def\epsfurx{#3}\global\def\epsfury{#4}\fi}%
%
%
\def\epsfsize#1#2{\epsfxsize}%
%
%

\epsfverbosetrue                        
\abovefigskip=\baselineskip             
\belowfigskip=\baselineskip             
\global\let\figtitlefont\bf             
\global\figtitleskip=.5\baselineskip    

\font\tenmsb=msbm10   
\font\sevenmsb=msbm7
\font\fivemsb=msbm5
\newfam\msbfam
\textfont\msbfam=\tenmsb
\scriptfont\msbfam=\sevenmsb
\scriptscriptfont\msbfam=\fivemsb
\def\Bbb#1{\fam\msbfam\relax#1}
\let\nd\noindent 

\def\natural{{\rm I\kern-.18em N}}
\newskip\ttglue


\def\eightpoint{\def\rm{\fam0\eightrm}  
  \textfont0=\eightrm \scriptfont0=\sixrm \scriptscriptfont0=\fiverm
  \textfont1=\eighti  \scriptfont1=\sixi  \scriptscriptfont1=\fivei
  \textfont2=\eightsy  \scriptfont2=\sixsy  \scriptscriptfont2=\fivesy
  \textfont3=\tenex  \scriptfont3=\tenex  \scriptscriptfont3=\tenex
  \textfont\itfam=\eightit  \def\it{\fam\itfam\eightit}
  \textfont\slfam=\eightsl  \def\sl{\fam\slfam\eightsl}
  \textfont\ttfam=\eighttt  \def\tt{\fam\ttfam\eighttt}
  \textfont\bffam=\eightbf  \scriptfont\bffam=\sixbf
    \scriptscriptfont\bffam=\fivebf  \def\bf{\fam\bffam\eightbf}
  \tt  \ttglue=.5em plus.25em minus.15em
  \normalbaselineskip=9pt
  \setbox\strutbox=\hbox{\vrule height7pt depth2pt width0pt}
  \let\sc=\sixrm  \let\big=\eightbig \normalbaselines\rm}

\font\eightrm=cmr8 \font\sixrm=cmr6 \font\fiverm=cmr5
\font\eighti=cmmi8  \font\sixi=cmmi6   \font\fivei=cmmi5
\font\eightsy=cmsy8  \font\sixsy=cmsy6 \font\fivesy=cmsy5
\font\eightit=cmti8  \font\eightsl=cmsl8  \font\eighttt=cmtt8
\font\eightbf=cmbx8  \font\sixbf=cmbx6 \font\fivebf=cmbx5

\def\eightbig#1{{\hbox{$\textfont0=\ninerm\textfont2=\ninesy
        \left#1\vbox to6.5pt{}\right.\enspace$}}}

\def\ninepoint{\def\rm{\fam0\ninerm}  
  \textfont0=\ninerm \scriptfont0=\sixrm \scriptscriptfont0=\fiverm
  \textfont1=\ninei  \scriptfont1=\sixi  \scriptscriptfont1=\fivei
  \textfont2=\ninesy  \scriptfont2=\sixsy  \scriptscriptfont2=\fivesy
  \textfont3=\tenex  \scriptfont3=\tenex  \scriptscriptfont3=\tenex
  \textfont\itfam=\nineit  \def\it{\fam\itfam\nineit}
  \textfont\slfam=\ninesl  \def\sl{\fam\slfam\ninesl}
  \textfont\ttfam=\ninett  \def\tt{\fam\ttfam\ninett}
  \textfont\bffam=\ninebf  \scriptfont\bffam=\sixbf
    \scriptscriptfont\bffam=\fivebf  \def\bf{\fam\bffam\ninebf}
  \tt  \ttglue=.5em plus.25em minus.15em
  \normalbaselineskip=11pt
  \setbox\strutbox=\hbox{\vrule height8pt depth3pt width0pt}
  \let\sc=\sevenrm  \let\big=\ninebig \normalbaselines\rm}

\font\ninerm=cmr9 \font\sixrm=cmr6 \font\fiverm=cmr5
\font\ninei=cmmi9  \font\sixi=cmmi6   \font\fivei=cmmi5
\font\ninesy=cmsy9  \font\sixsy=cmsy6 \font\fivesy=cmsy5
\font\nineit=cmti9  \font\ninesl=cmsl9  \font\ninett=cmtt9
\font\ninebf=cmbx9  \font\sixbf=cmbx6 \font\fivebf=cmbx5
\def\ninebig#1{{\hbox{$\textfont0=\tenrm\textfont2=\tensy
        \left#1\vbox to7.25pt{}\right.$}}}

\def\Z{{\Bbb Z}}
\def\chix{{\raise.5ex\hbox{$\chi$}}}
\def\chixa{{\chix\lower.2em\hbox{$_A$}}}

\def\real{{\rm I\kern-.2em R}}
\def\integer{{\rm Z\kern-.32em Z}}
\def\complex{\kern.1em{\raise.47ex\hbox{
            $\scriptscriptstyle |$}}\kern-.40em{\rm C}}
\def\vs#1 {\vskip#1truein}
\def\hs#1 {\hskip#1truein}
  \hsize=6.2truein \hoffset=.23truein 
  \vsize=8.8truein 
\pageno=1 \baselineskip=12pt
  \parskip=0 pt \parindent=20pt 
\overfullrule=0pt \lineskip=0pt \lineskiplimit=0pt
  \hbadness=10000 \vbadness=10000 
     \pageno=0
     
     \footline{\ifnum\pageno=0\hss\else\hss\tenrm\folio\hss\fi}
     \hbox{}
     \vskip 1truein\centerline{{\bf First Order Phase Transition of a Long Polymer Chain}}
     \vskip .2truein\centerline{by}
     \vskip .2truein
\centerline{{David Aristoff}
\ \ and\ \  {Charles Radin}
\footnote{*}{Research supported in part by NSF Grant DMS-0700120\hfil}}

\vskip .1truein
\centerline{ Mathematics Department, University of Texas, Austin, TX 78712} 
\vs.5 \centerline{{\bf Abstract}} 
\vs.2 \nd
We consider a model consisting of a self-avoiding polygon 
occupying a variable density of the sites of a square lattice. 
A fixed energy is associated with each $90^\circ$-bend of the polygon.
We use a grand canonical ensemble, introducing parameters $\mu$ and $\beta$ 
to control average density
and average (total) energy of the polygon, and show by Monte Carlo
simulation that the model has a first order, nematic 
phase transition across 
a curve in the $\beta$-$\mu$ plane. 

\vs2
\centerline{November, 2010}
\vs1
\centerline{PACS Classification:\ \ 82.35.Lr, 64.70.M-, 64.60.De, 36.20.Fz} 

     \vfill\eject
\nd
{\bf 1. Introduction}
\vs.1

In polymer physics self-avoiding walks have been used for many years
to model long chain molecules [1,2]. One of
the oldest such models, due to Flory [3], consists of a single
self-avoiding random polygon occupying {\it all} the sites of a square
lattice, with the randomness controlled by the 
total energy associated with $90^\circ$-bends in the polygon. Thinking
of the polygon as made of many flexibly connected monomers, the model is a canonical
ensemble with the temperature behavior analyzed only at the optimally
high particle density of 1. At high
temperature the Flory model behaves like a disordered fluid, while at low
temperature the model displays long-range nematic order. (Flory was
modeling the ``melting'' of a nematically ordered polymer [3].) Although it is
generally accepted that there is a true phase transition between these
regimes in the model, there has been a dispute over the character of
the transition [4], which Flory had originally predicted
to be first order.
For the Flory model the controversy seems to have been settled by a
recent paper [5] of Jacobsen and Kondev which shows that the transition
is second order. (There is a useful summary of the history of the
controversy in [5].)
They also suggest, however, that the transition may become first order 
at lower density if vacancies are permitted (see Section VII C in [5]).
In our paper we concentrate on this generalization of the Flory model which
allows for vacant lattice sites, so that our random polygon is now
controlled by two parameters, temperature and
chemical potential. We show by Monte Carlo simulation that the 
model has a first order phase transition. We 
also use order parameters to illustrate the nematic nature of the transition.

We note some other variations and applications of the Flory model. One
variation uses an ensemble of polygons rather than one long
polygon. Although the connection between the two models is
unclear, it is of interest that there have been conflicting results on
the character of the transition in this case
also [6,7]. In a different direction we note that variants of the Flory model 
have been applied to
nonequilibrium materials, specifically to crumpled or confined sheets
(in 3 dimensions) and wires (in 2 or 3 dimensions). For instance in
[8] we modeled a progressively confined wire in 2 dimensions by a
version of the model used here, but on a triangular lattice, showing a
similar transition. A field theoretic (continuum) model of a
variably confined wire
by Bou\'e and Katsev [9], however, found a second order
transition, and it remains to understand the origins of this
difference between their continuum model and our lattice models.

\vs.2
\nd {\bf 2. The Model, and Results}
\vs.1

Consider the set $\cal W$ of all self-avoiding polygons (i.e., 
closed self-avoiding loops) on the square lattice with 
periodic boundary conditions, $L = ({\Z}/ v {\Z})^2$,
with $v$ a fixed positive integer. 
The energy $E(w)$ of polygon $w$ in $\cal W$ is defined as the number of right angles 
in $w$, and the length $N(w)$ of $w$ is the number of unit line segments 
in $w$. Given an inverse temperature $\beta = {1}/{T}$ and a chemical potential 
$\mu$, the free energy of the model is $\beta E -\beta \mu N - S(E,N)$ where $S(E,N)$
is the entropy, that is, the natural logarithm of the volume in phase space of self-avoiding polygons at 
fixed $E$ and $N$. 

\epsfig 1\hsize; 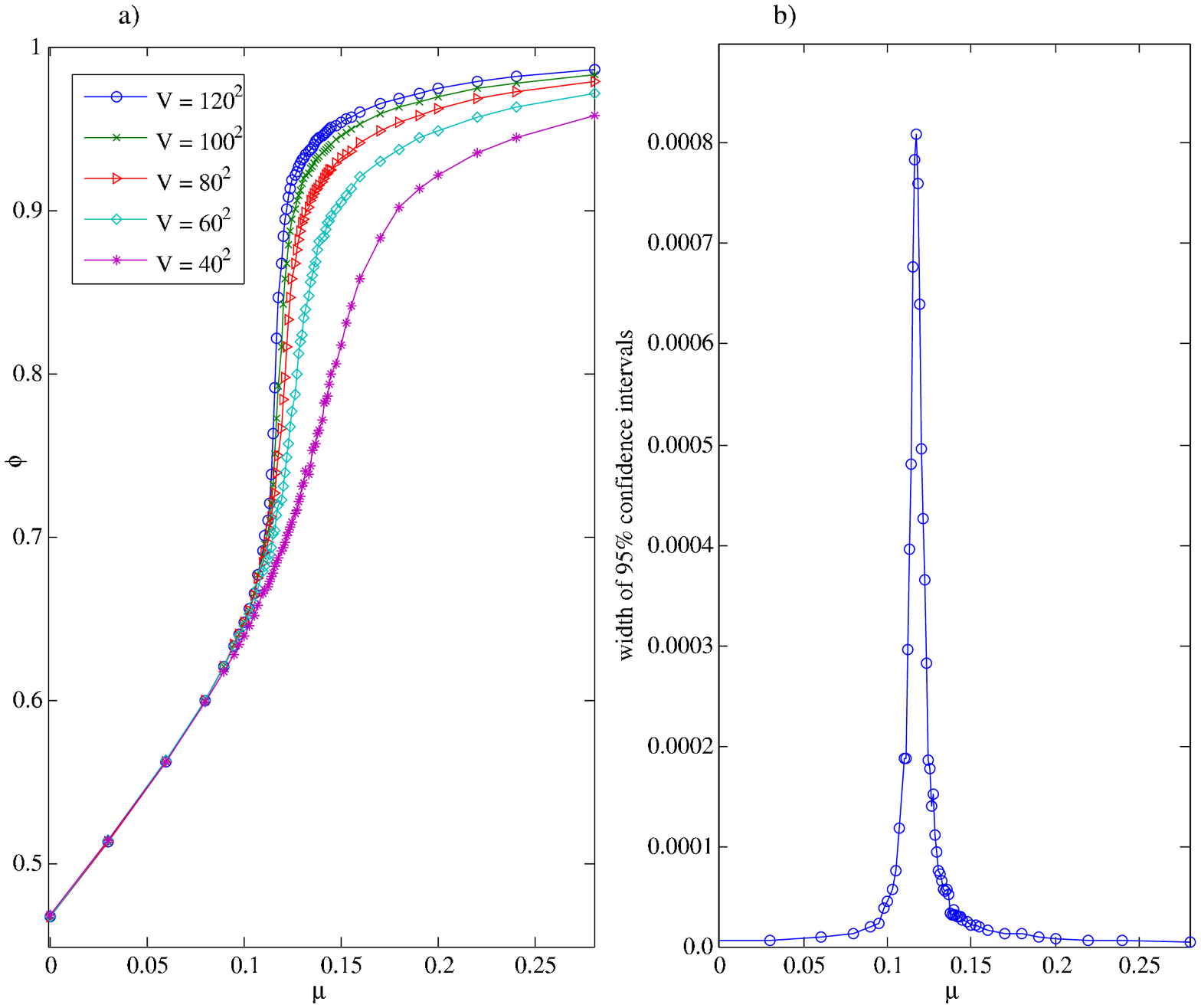;

\nd {\bf Figure 1.} {a) The graph of average density vs.\ $\mu$ at $\beta = 1.5$, for system volumes $V = 40^2$ through $V = 120^2$.
 b) Width of $95\%$ confidence intervals for average density, for $V = 120^2$.}
\vs.2
As usual in a grand canonical ensemble this 
is optimized by the probability measure $m_{\beta,\mu}$ defined on the subsets of $\cal W$ by 
$$m_{\beta,\mu}(w) = {{1}\over{Z_{\beta,\mu}}}e^{-\beta(E(w)-\mu N(w))},\eqno{1)}$$
for $w \in \cal W$, where $Z_{\beta,\mu}$ is the appropriate normalization. In this
notation we have suppressed 
the dependence of $m_{\beta,\mu}$ and $Z_{\beta,\mu}$ on the system volume $V = v^2$. 

To simulate the model we fix either $\beta$ or $\mu$, and then 
slowly increase the other parameter, starting from well into the
disordered regime. The basic Monte Carlo step is as follows 
(see pgs. 41-44 in [10]). 
Given a polygon $w(t)$ at step $t$ in the 
simulation, we introduce, with probability $p_i$, 
a trial configuration $w(t)'$ which changes 
the length of $w(t)$ by $\ell_i$. If $w(t)'$ is not self-avoiding 
then we take $w(t+1) = w(t)$; otherwise we set $w(t+1) = w(t)'$ 
with probability $q = \min(Q,1)$, and $w(t+1) = w(t)$ with probability $1-q$, where 
$$Q = e^{\beta[\mu\ell_i+E(w(t))-E(w(t)')]}\eqno{2)}$$
Here $p_1 = p_2 = 2/5$, $p_3 = 1/5$, $\ell_1 = -2$, $\ell_2 = 2$ and $\ell_3 = 0$. 
\epsfig 1\hsize; 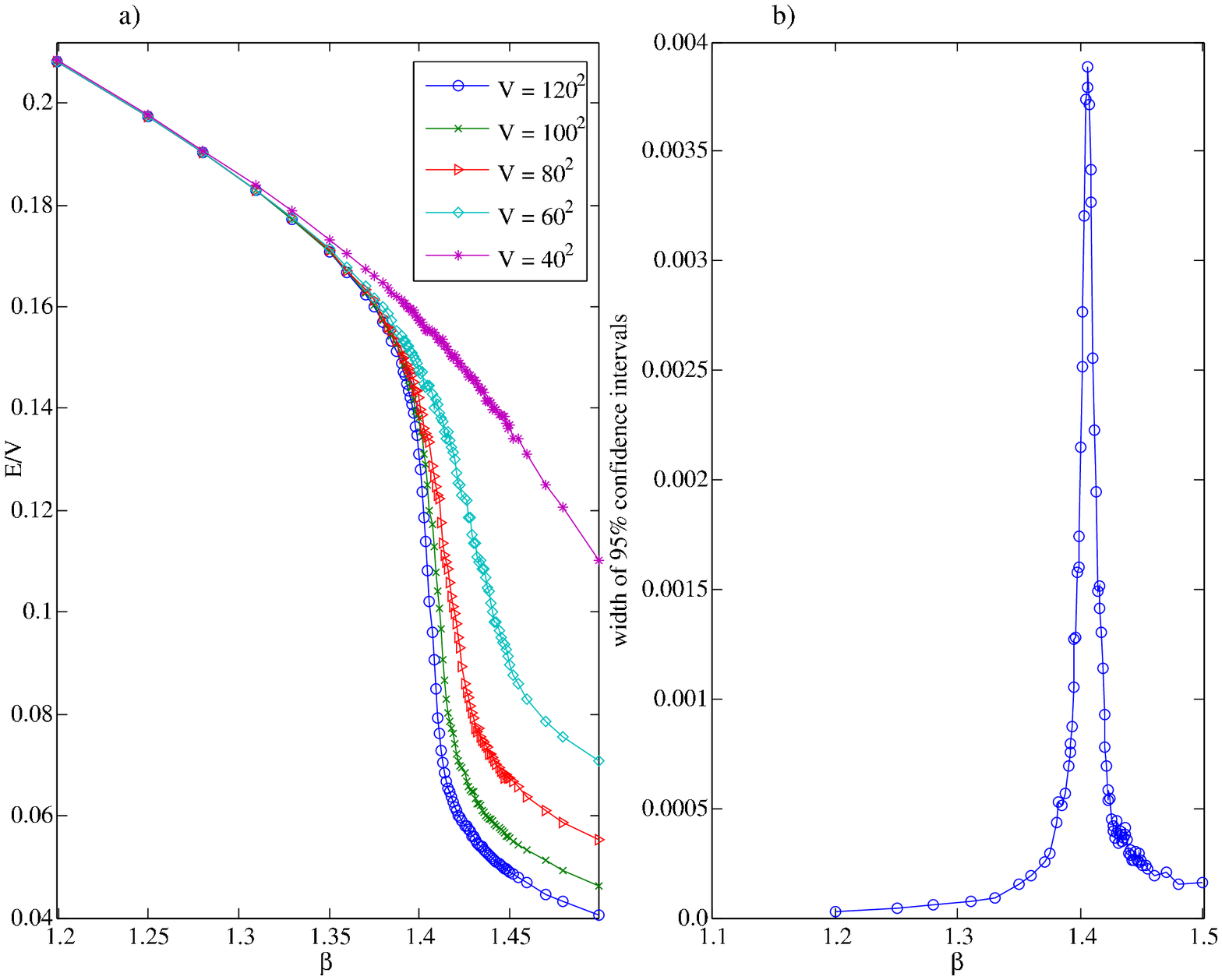;
\nd {\bf Figure 2.} {a) The graph of average energy per volume vs.\ $\beta$ at $\mu = 0.15$, for system volumes $V = 40^2$ through $V = 120^2$. 
b) Width of $95\%$ confidence intervals for average energy per volume, for $V = 120^2$.}
\vs.2

To determine whether the simulation for each pair $(\beta,\mu)$ has sufficiently 
many Monte Carlo steps, we compute a ``mixing time'' as 
the smallest $t$ such that the standard autocorrelation function 
$${{1}\over {(n-t)\sigma^2}} \sum_{i=1}^{n-t} (meas(w_i)-\lambda)\cdot (meas(w_{i+t})-\lambda)\eqno{3)}$$
falls below zero. Here $meas$ represents any of our various measurements, described 
below, $\lambda$ and $\sigma^2$ are the sample average and variance (respectively) of $meas$ 
over the simulation of $(\beta,\mu)$, and $w_i$ is the $i$th configuration 
in the simulation of $(\beta,\mu)$, with $n$ total steps. 
We found that our simulations of each $(\beta,\mu)$ were, in the worst cases, 
at least 5 mixing times long (on average), and we therefore believe our Monte Carlo runs are 
reasonably close to sampling the distributions $m_{\beta,\mu}$. We repeated each 
of our simulations 100 times, and obtained 95$\%$-confidence
intervals for $meas$ from Student's $t$-distribution 
with 99 degrees of freedom on the average values of $meas$ over each simulation. 
(Measurements related to specific heat were calculated differently, and are discussed 
below.) A single simulation of the largest system ($V= 120^2$) contains $8\times 10^{11}$ 
basic Monte Carlo steps.

\epsfig 1\hsize; 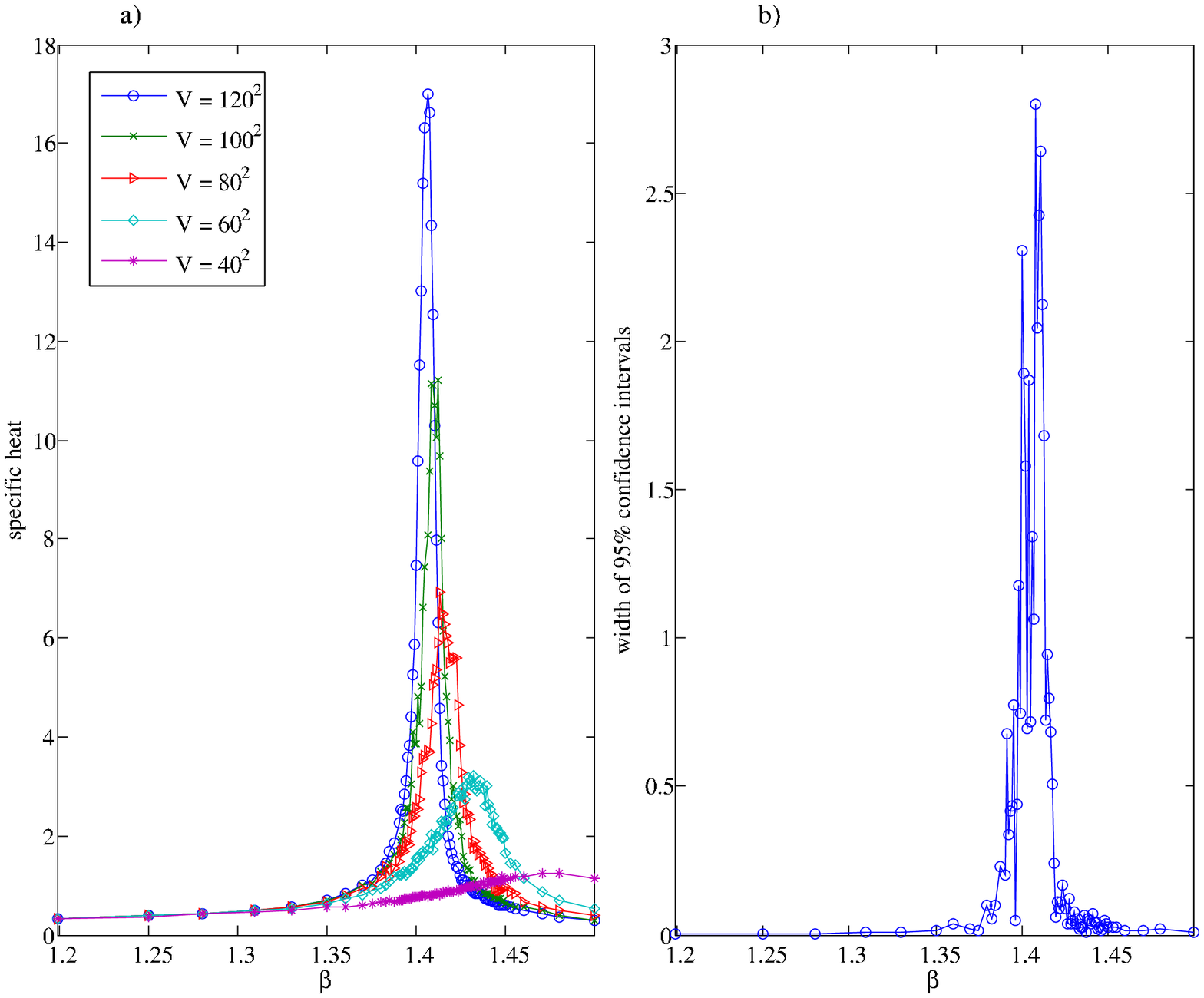;
\nd {\bf Figure 3.} {a) The graph of specific heat vs.\ $\beta$ at $\mu = 0.15$, for system volumes $V = 40^2$ through $V = 120^2$.  
b) Width of $95\%$ confidence intervals for specific heat, for $V = 120^2$.}
\vs.2

We measure average energy per volume ${\langle E\rangle_{\beta,\mu}}/{V}$, average density
$\langle \phi \rangle_{\beta,\mu}$, as well as order parameters $corr$ and $lay$, which were introduced in [8] and are 
defined as follows. Given a polygon $w$, $corr(w)$ is the proportion of 
edges in $w$ which have the same orientation 
(horizontal or vertical) as a randomly chosen edge in $w$. Given $w$, $lay(w)$ is 
the normalized volume ${u^2}/{V}$ of the largest square sublattice $L' = ({\Z}/u{\Z})^2 \subset L$ 
such that the orientation of $w$ (horizontal or vertical) at the origin 
agrees with the orientation of $w$ at $80\%$ or more of the sites in
$L'$. (We choose  
an orientation for the polygon $w$ so that each lattice site has a unique horizontal 
or vertical orientation.) 

\epsfig .8\hsize; 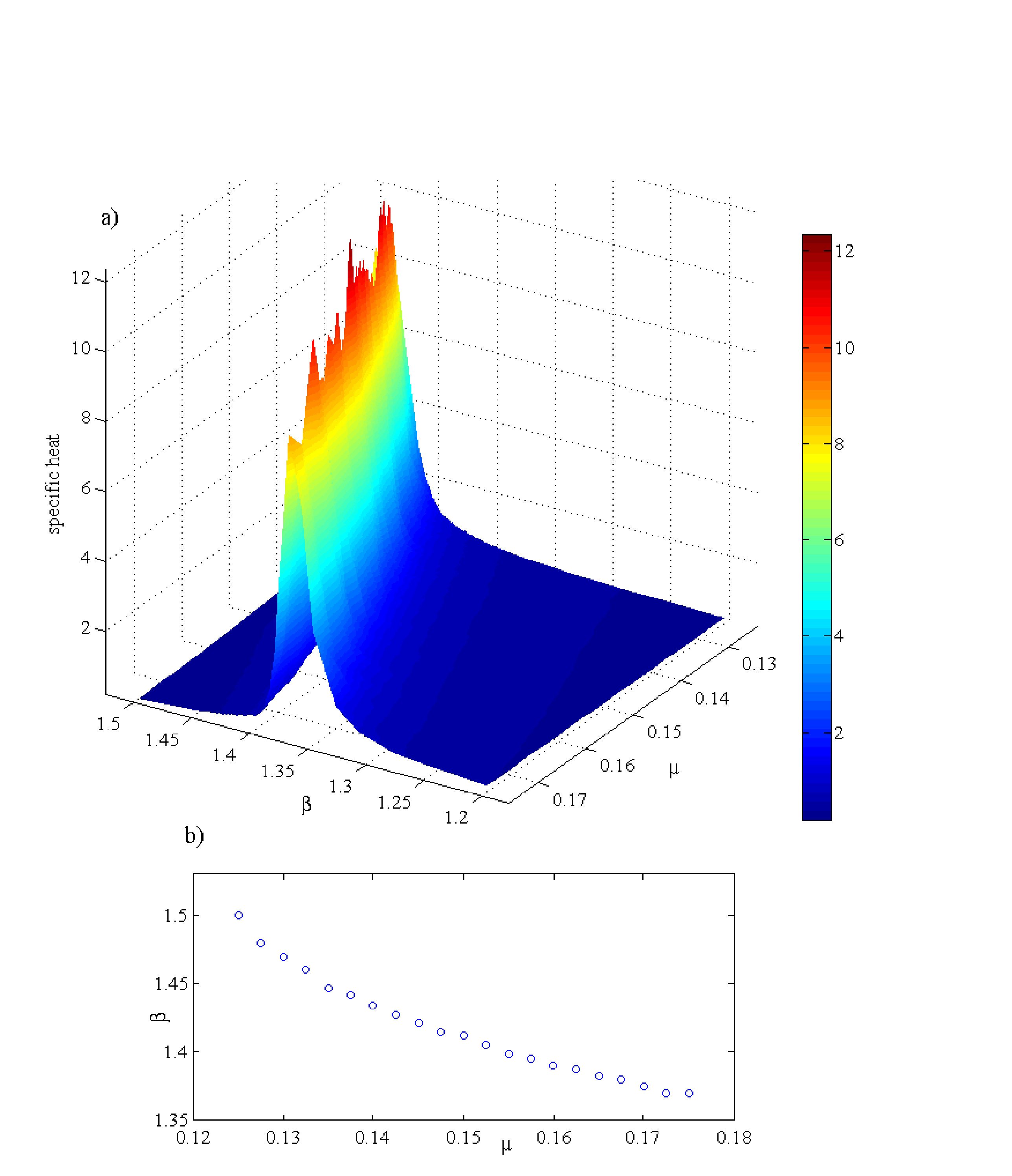;
\nd {\bf Figure 4.} {a) The graph of specific heat vs.\ $\beta$ and $\mu$, for $V = 100^2$.
b) Estimation of $\beta$ values which maximize specific heat at fixed $\mu$.}

\vs.2

We compute specific heat $({1}/{V}){\partial  \langle E\rangle_{\beta,\mu}}/{\partial T}$ from fluctuations, that is, 
$$T^2{{\partial  \langle E\rangle_{\beta,\mu}}\over {\partial T}} =
\langle E \rangle_{\beta,\mu} \langle \mu N - E
\rangle_{\beta,\mu}-\langle E\mu N-E^2 \rangle_{\beta,\mu}.\eqno{4)}$$
To compute the values of $\langle \cdot \rangle_{\beta,\mu}$ from equation (4), we took averages of 
the relevant measurements $meas$ over 100 independent simulations. Then for $95\%$-confidence 
intervals we repeated this process 4 times, and used the Student's $t$-distribution with 
$3$ degrees of freedom. We checked that the resulting curve agreed with 
the numerical derivative of energy. 

In contrast with [8] we simulate well into the ordered regime and find 
direct evidence of a first order phase transition. 
In particular the trends with increasing system size 
in the curves of Figs.\ 1 and 2 strongly suggest that 
the average density and average energy per volume both
develop jump discontinuities at the transition, in the infinite volume limit. 
In confirmation, Fig.\ 3 shows the specific heat developing a delta
function singularity at the transition.

\epsfig 1\hsize; 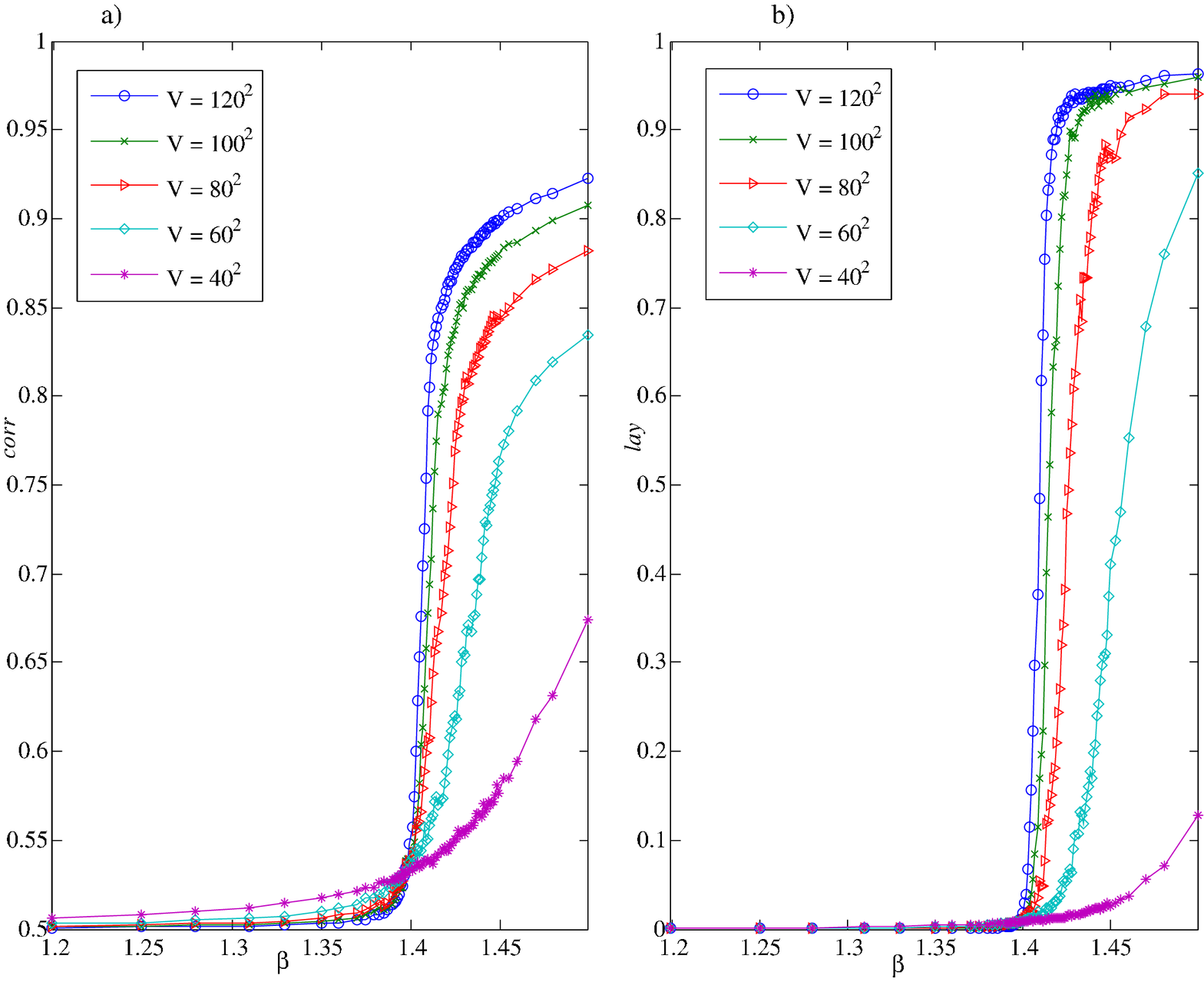;
\nd {\bf Figure 5.} {a) The graph of {\it corr} vs.\ $\beta$ at $\mu = 0.15$, for volumes $V = 40^2$ to $V = 120^2$. 
  b) The graph of {\it lay} vs.\ $\beta$ at $\mu = 0.15$, for volumes $V = 40^2$ to $V = 120^2$.}
\vs.2
We plot the specific heat surface as a function of $\beta$ and $\mu$,
as well as the $(\beta,\mu)$-coordinates of the maximum of specific heat, at various
$0.125 \le \mu \le 0.175$, $1.2 \le \beta \le 1.5$ in Fig.\ 4. The
latter gives an indication
of the transition curve;
note that as $\mu$ increases, the temperature at which the transition
occurs increases.

As evidence of an nematic transition, the measurements $corr$
and $lay$ (see Figs.\ 5 and 6) 
exhibit a jump discontinuity at the transition from their
disorder values of ${1}/{2}$ and zero, respectively.


\epsfig 1\hsize; 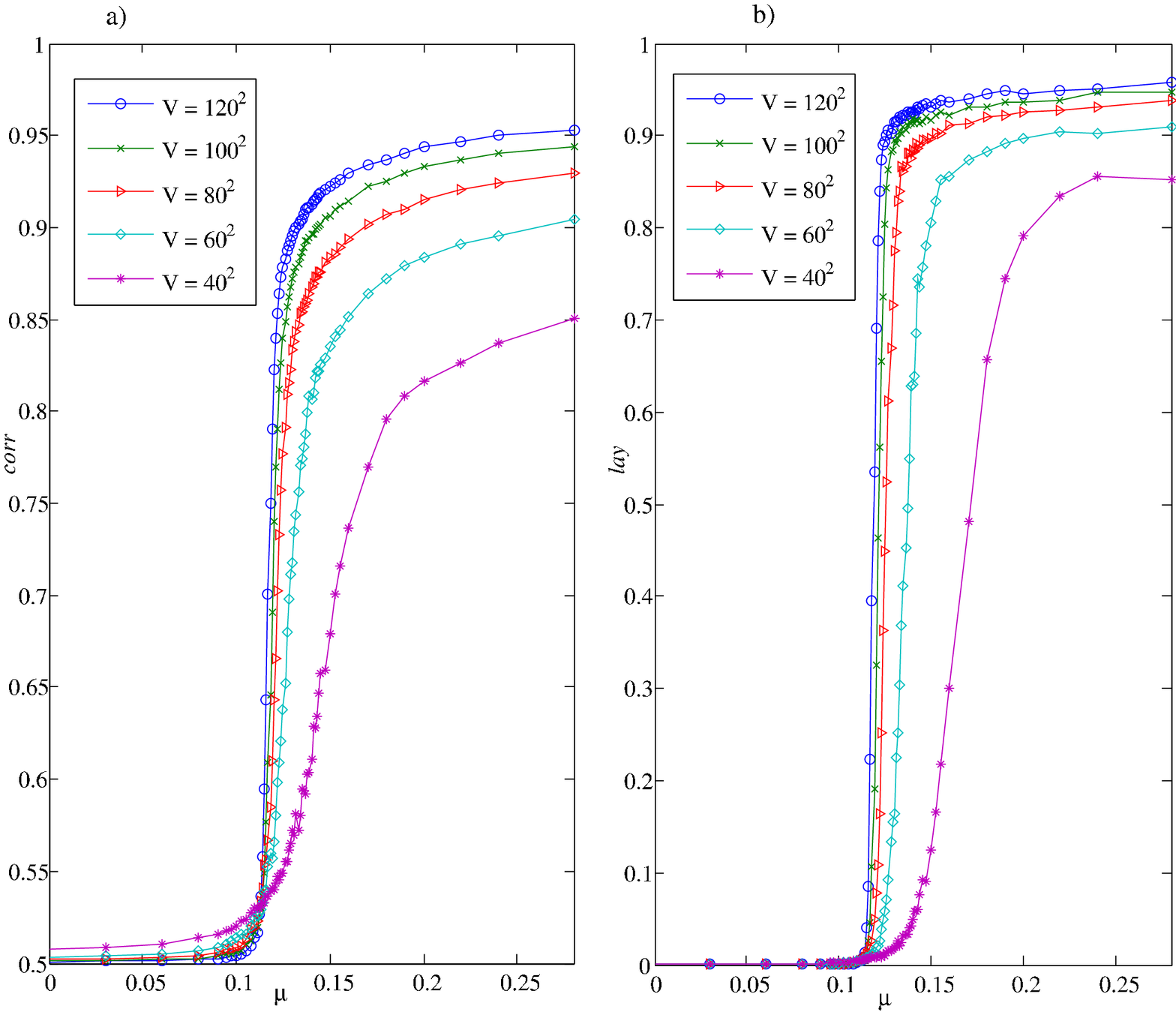;
\nd {\bf Figure 6.} {a) The graph of {\it corr} vs.\ $\mu$ at $\beta = 1.5$, for volumes $V = 40^2$ to $V = 120^2$. 
b) The graph of {\it lay} vs.\ $\mu$ at $\beta = 1.5$, for volumes $V = 40^2$ to $V = 120^2$.}
\vs.2

\nd {\bf 3. Nonequilibrium}
\vs.1
We mentioned above that with variable density added, a version of the Flory
model on a triangular lattice
has
been used [8] to model wires progressively confined in 2
dimensions. (See [11] for a related approach, and further references.)
We add here a note on the modeling of such
nonequilibrium materials. Since the polygons are self-avoiding, these
lattice models introduce a unit length scale for the width of the
wire. Now if it requires energy $E$ to bend one wire of unit
thickness to a given radius of curvature it would require $mE$ to bend
a loose bundle of $m$ parallel wires, but because of the
interconnectedness it would require more than
$mE$ to bend a single wire of thickness $m$. Therefore the bending
energy is highly nonlinear in the thickness of the wire, growing
faster than the square of the thickness, and so energy is an
independent parameter in our modeling.

\epsfig 1\hsize; 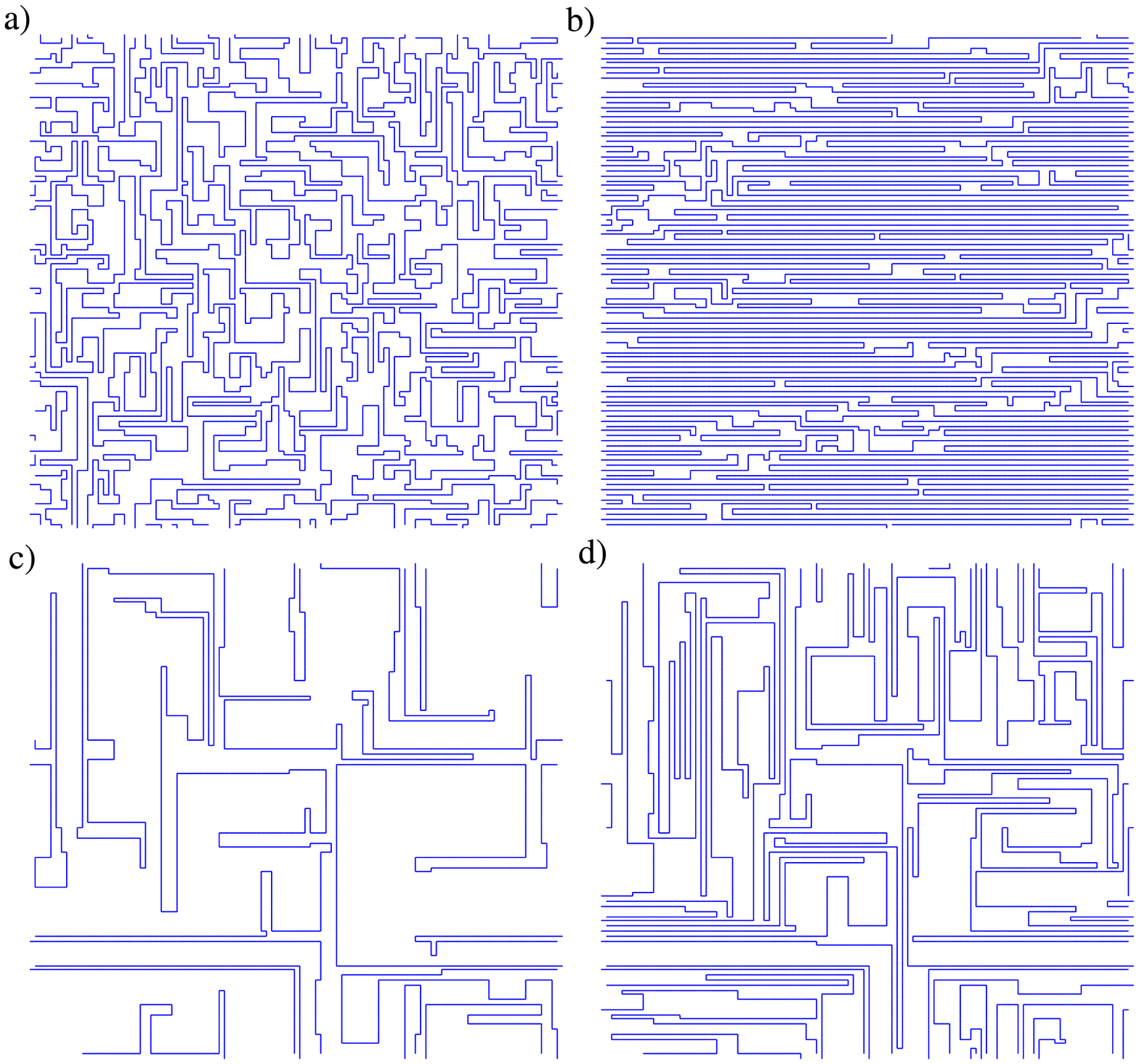;

\nd {\bf Figure 7.} {Configurations with $V = 100^2$ in equilibrium at:  

a) $(\beta, \mu) = (1.5,0.03)$ and density $\phi = 0.392$

b) $(\beta,\mu) = (1.5, 0.13)$ and density $\phi = 0.655$, 

c) $(\beta,\mu) = (2.5,0.01)$ and density $\phi = 0.146$,

d) $(\beta,\mu) = (2.5,0.025)$ and density $\phi = 0.392$.}
\vs.2
Fig. 7\ shows typical polygons 
with low and high volume fraction, but at two very different temperatures. 
This could be useful in estimating whether experimentally confined
materials are in an ordered or disordered regime. An alternative
method would be to compute order parameters such as $lay$ or $corr$.

\vs.2 \nd

\vs.2 \nd
{\bf 4. Conclusion} 
\vs.1

We have introduced a version of the Flory model that allows for a
positive fraction of vacancies, and shown by Monte Carlo simulation
that the model has a first order, nematic phase
transition. In terms of the (inverse) temperature $\beta$ and chemical
potential $\mu$ of our grand canonical ensemble, we find that the
transition lies approximately on the curve shown in Fig.\ 4 b).

We conclude by contrasting the results in this paper with those in
[8], which uses a very similar model, but on a triangular lattice and
therefore with greater complexity of the interaction energies. A grand
canonical ensemble was also used in [8], but simulations were confined
to a single isotherm. 

The evidence in [8] of a nematic
transition is based on two order parameters, {\it lay} and {\it corr},
used also in this paper. The arguments in [8] are based heavily on
trends in these order parameters as the system size is increased,
namely that at low chemical potential $\mu$ both order parameters decrease
monotonically toward their disordered values, while at higher $\mu$
both order parameters increase monotonically away from their disordered
values. This strongly suggests a nematic
transition, which would be expected to be first order. The simulations in [8] became
unreliable moving into the ordered regime, so in particular no direct
evidence was given of a discontinuity of a first derivative of the free energy,
namely volume fraction or energy density.


For the present paper, which also reports a nematic transition in a
model similar to [8] but on a square lattice, as in the original Flory
model [4,5], we were able to make reliable simulations well into the
ordered regime, and simulated on a grid of values of $\mu$ and
$\beta$. The main improvement over [8] is that now we are able to show discontinuities in volume
fraction and energy density, the usual hallmarks of a first
order transition, as well as discontinuities in {\it corr} and {\it
  lay}, which clarify the nematic nature of the transition. 
We feel this is strong, direct evidence supporting the suggestion in [5]
that the second order transition which they find at density 1 in the
Flory model becomes first order at lower density when vacancies 
are included in the model. It still remains to reconcile this behavior
with that found in [9] in their continuum model of confined loops.


\bigskip

\vs.2
\nd {\bf Acknowledgements.}\ 
We are grateful to E.\ Katzev, S.\ Deboeuf, A.\ Boudaoud and N.\ Menon
for useful discussions.
\vfill \eject
\centerline{{\bf References}}
\vs.2
\item{1} P.J. Flory, {\it Statistical Mechanics of Chain Molecules},
Wiley, 1969.

\item{2} P.-G. de Gennes, {\it Scaling Concepts in Polymer Physics},
Cornell University Press, Ithaca, 1979.

\item{3} P. J. Flory, Proc. R. Soc. London, Ser. A 234, 60 (1956).

\item{4} G. I. Menon and R. Pandit, Phys. Rev. E 59, 787 (1999).

\item{5} J. L. Jacobsen and J. Kondev, Phys. Rev. E, 69 (2004) 066108.

\item{6} J. F. Nagle, Proc. R. Soc. London, Ser. A 337, 569 (1974).

\item{7} A. Baumgartner and D. Yoon, J. Chem. Phys. 79, 521 (1983).

\item{8} D. Aristoff and C. Radin, Europhys. Lett. 91 (2010) 56003.

\item{9} L. Bou\'e  and E. Katzav, Europhys. Lett. 80 (2007) 54002.

\item{10} E.J. Janse van Rensburg, J. Phys. A 42 (2009) 323001.

\item{11} M. Adda-Bedia, A. Boudaoud, L. Bou\'e and S. Deboeuf,
  arXiv:1009.1001v1.



\vfill
\end